\documentstyle[12pt,aaspp4,flushrt,psfig]{article}

\newcommand{\kms}{\hbox{${\rm km\,s^{-1}}$}}
\newcommand{\ea}{{\it et al.}}
\newcommand{\vi}{\hbox{$V\!-\!I$}}
\newcommand{\feh}{\hbox{${\rm [Fe/H]}$}}
\newcommand{\teff}{\hbox{${\rm T_{eff}}$}}
\newcommand{\logl}{\hbox{$(L/L_{\odot})$}}
\newcommand{\dmv}{\hbox{$(m\!-\!M)_{V}$}}

\newcommand{\msun}{\hbox {${\rm M}_{\odot}$}}
\newcommand{\dyz}{\hbox {$\Delta Y/\Delta Z$}}
\newcommand{\ebv}{\hbox {${\rm E(B\!-\!V)}$}}
\tighten

\slugcomment{\bf FINAL DRAFT }

\lefthead{Chaboyer, Green, \& Liebert}
\righthead{The Old, Metal-Rich Open Cluster NGC\,6791}

\begin{document}
\tighten

\title{ THE AGE, EXTINCTION AND DISTANCE OF THE OLD, METAL-RICH OPEN CLUSTER NGC\,6791}

\author{Brian Chaboyer\altaffilmark{1,2}, Elizabeth M. Green, and James Liebert}
\affil{Steward Observatory, University of Arizona, Tucson AZ 85721} 
\affil{chaboyer@as.arizona.edu, bgreen@as.arizona.edu,
liebert@as.arizona.edu}
\altaffiltext{1}{Hubble Fellow}
\altaffiltext{2}{Current Address: Department of Physics and Astronomy,
6127 Wilder Lab, Dartmouth College, Hanover, NH  03755-3528}


\author{}
\affil{} 
\affil{}

\begin{abstract}

An extensive grid of metal-rich isochrones utilizing the latest
available input physics has been calculated for comparison with the
old, metal-rich open cluster NGC\,6791.  The isochrones have been
simultaneously fit to BV and VI color magnitude diagrams, with the
same composition, reddening and distance modulus required for both
colors.  Our best fitting isochrone assumes $\feh = +0.4$, scaled
solar abundance ratios, and $\dyz = 2$ ($Y = 0.31$), yielding an
excellent fit to the data at all points along the major sequences.
The resulting age is 8 Gyr, with $\ebv = 0.10$ and $\dmv = 13.42$.
The derived cluster parameters are fairly robust to variations in the
isochrone \feh\ and helium abundances.  All of the acceptable fits
indicate that $0.08 \le \ebv \le 0.13$, $13.30 \le \dmv \le 13.45$,
and that NGC\,6791 has an age of $8.0\pm 0.5\,$Gyr.  The fits also
suggest that $\dyz$ lies between 1 and 3.  A metallicity as low as
solar is clearly ruled out, as is $\dyz = 0$.  Comparison with
previous isochrone studies indicates that the derived reddening is
primarily due to our use of the most recent color transformations,
whereas the age depends upon both the colors and the input physics.
Our isochrones provide an excellent fit to the Hyades zero-age main
sequence as determined by Hipparcos, providing evidence that our
derived reddening and distance modulus are reliable.

\end{abstract}

\vfill
\eject

\section{Introduction}

The old open cluster NGC\,6791 is a critically important object for
the study of stellar evolution, the galactic age-metallicity relation,
and population synthesis.  It is one of the two or three oldest of the
known open clusters (Phelps, Janes, \& Montgomery 1994), yet it has
the highest metallicity (Friel \& Janes 1993; Garnavich \ea\ 1994;
Peterson \& Green 1998), making it a unique local example of the kind
of old, metal-rich population expected in the centers of elliptical
galaxies and spiral bulges.  It is also one of the most populous of
all the open clusters (Ka\l u\.zny \& Udalski 1992).  As a result,
NGC\,6791 has been more intensively studied than any other old disk
cluster except M67.  Numerous color magnitude diagram (CMD)
investigations (Kinman 1965; Anthony-Twarog \& Twarog 1985; Ka\l
u\.zny 1990; Ka\l u\.zny \& Udalski 1992; Ka\l u\.zny \& Rucinski
1993; Montgomery, Janes, \& Phelps 1994; Garnavich \ea\ 1994; Ka\l
u\.zny \& Rucinski 1995), isochrone studies (Demarque, Green, \&
Guenther 1992; Meynet \ea\ 1993; Carraro \ea\ 1994; Tripicco \ea\
1995) and other photometric and spectroscopic investigations {\it
e.g.}\ (Harris \& Canterna 1981; Liebert, Saffer, \& Green 1994;
Twarog, Ashman, \& Anthony-Twarog 1997; Eggen 1998) have attempted to
constrain the reddening, distance modulus, and age, with often
contradictory results.  Derived ages have ranged as low as 5.5 and as
high as 12.5 Gyr, given the uncertainties in the reddening ($0.09 <
\ebv < 0.26$) and the composition (both the metallicity and the
assumed helium abundance).  Equally important, the use of a wide
variety of theoretical isochrones and different transformations from
theoretical \teff\ and \logl\ to observed CMD's have made it very
difficult to compare the results obtained by various investigators
(Friel 1995).

Reliable ages, extinction and distances from isochrone fitting depend
on several factors.  First, the observational CMD's must be of very
high quality: deep, photometrically accurate, well populated along the
major sequences, and reasonably free of confusion from field stars or
binaries.  Second, the stellar composition must be known fairly
precisely, since any errors will result in compensating variations in
\ebv, \dmv, and age.  Third, the effects of uncertainties in the
isochrone input physics should be small compared to the desired
accuracy of the cluster parameters to be fit.  While this is now
believed to be an achievable goal, it is difficult to estimate the
true uncertainties in the isochrone input physics.  Nevertheless, it
is important to note that even relatively large uncertainties, such as
the accuracy of the opacity tables and the choice of mixing length,
have only small effects on the derived ages, reddenings and distance
moduli, as long as the isochrone fits are made to the main sequence
and turnoff regions, and the models are properly calibrated to a
standard solar model (Demarque, Green, \& Guenther 1992).  Finally,
the theoretical quantities must be accurately transformed into the
observational plane.  This is especially important for high $Z$
models, since metal-rich stars have high line blanketing which must be
correctly included in the synthetic stellar atmospheres.

The observational data are very good for NGC\,6791.  Multiple, very
precise CMD's are available in several colors (see the references
above).  Preliminary proper motion membership probabilities from
Cudworth's (1994) ongoing survey show a high degree of separation
between cluster and field stars.  As discussed in \S\ref{composition},
the metallicity question is now much closer to resolution following
the first high dispersion abundance analysis of a cluster member
(Peterson \& Green 1998).  We have chosen to further constrain the
possible range of isochrone solutions by doing simultaneous fits in
$V$ {\it vs.}\ \bv\ and $V$ {\it vs.}\ \vi, requiring that the
distance modulus and reddening be the same in both CMD's.

New theoretical isochrones have been computed (\S \ref{sectmodel})
to match the observational data.  Recent improvements include
the use of OPAL opacities, diffusion, and color transformations based
on Kurucz's new models.  In a major advance, Kurucz (1993) has
published a new grid of synthetic model atmospheres with stellar
abundances as high as six times solar metallicity, based on a
comprehensive compilation of new atomic lines and molecular data.
Bessell, Castelli, \& Plez (1998) have tested colors based on these
atmospheres (and other similar grids now being completed) against
fundamental color-temperature relations derived from several sources.
They conclude that Kurucz's new models fit stars hotter than 4250K
very well for all UBVRIJKL indices, at least for solar abundances.  As
we will demonstrate in \S \ref{previous}, the color transformation is
undoubtedly one of the most important, and least discussed,
uncertainties involved in isochrone fitting.

In \S \ref{sectfits}, we examine our isochrone fits to NGC\,6791 for a
range of \feh, \dyz, and age.  We investigate the validity of our
results (\S \ref{compare}) by comparing our isochrones with 1) those
used in previous investigations of NGC\,6791, 2) Hipparcos data for
the moderately metal-rich Hyades ZAMS, and 3) BV and VI CMD's for the
canonical solar metallicity cluster M67.

\section{The Composition of NGC\,6791 \label{composition}}

Composition is one of the most crucial isochrone input parameters.
Unfortunately, the brightest stars in NGC\,6791 are all very cool,
heavily-blanketed giants.  Though there is widespread agreement that
their metallicity must be greater than solar, until recently there
were only two spectroscopic determinations of [Fe/H], both from
moderate resolution spectra.  Friel \& Janes (1993) used calibrated
spectroscopic indices (Friel 1987) from nine giants in NGC\,6791 to
derive $[Fe/H] = +0.19\pm \,0.19$, while Garnavich {\it et al.}\ found
$[Fe/H] \approx +0.22$ from the Ca~II infrared triplet lines in eight
giants.  Other investigators based their conclusions on calibrated
photometric indices or isochrone fits to the color magnitude diagram.

A much more accurate metallicity has now been derived from a high
resolution echelle spectrum of a probable blue horizontal branch star
in NGC\,6791, 2-17 (Peterson \& Green 1998).  This star is undoubtedly
a cluster member, as it has a very strong-lined spectrum, 93\% proper
motion membership probability (Cudworth 1994), a radial velocity
within 2~\kms\ of the cluster mean, and a surface gravity appropriate
to its position in the CMD.  The combination of high temperature
($7300K \pm \,50$) and high metallicity in 2-17 lead to line strengths
that are close to solar, reducing several sources of systematic bias.
Peterson \& Green find $[Fe/H] = +0.4\pm \,0.1$, with nearly solar
ratios of C, O, Ca, and Al, but enhancements of 0.2--0.3 dex in Mg,
Si, and Na, and +0.5 in N.

A very important question is whether such high abundances should be
considered representative of the cluster as a whole.  Peterson \&
Green considered 2-17 to be a horizontal branch star, but could not
rule out its being a very luminous, slowly rotating blue straggler.
In either case, the values for iron and other heavy elements should
reflect the primordial abundances, though C, N, and O, and possibly
also Na, Mg, and Al ({\it e.g.}\ Kraft 1994), may have been altered by
mixing.  The \feh\ of 2-17 is not likely to be unusual for NGC\,6791, given
the evidence for star-to-star uniformity among the heavy element
abundances found by Hufnagel, Smith, \& Janes (1995).  Their
spectroscopic indices for 29 red horizontal branch and giant branch
members of NGC\,6791 showed no indication of significant variations in
Fe or Ca, though there was evidence for some variation in CN band
strengths.

\begin{figure}
\centerline{\psfig{figure=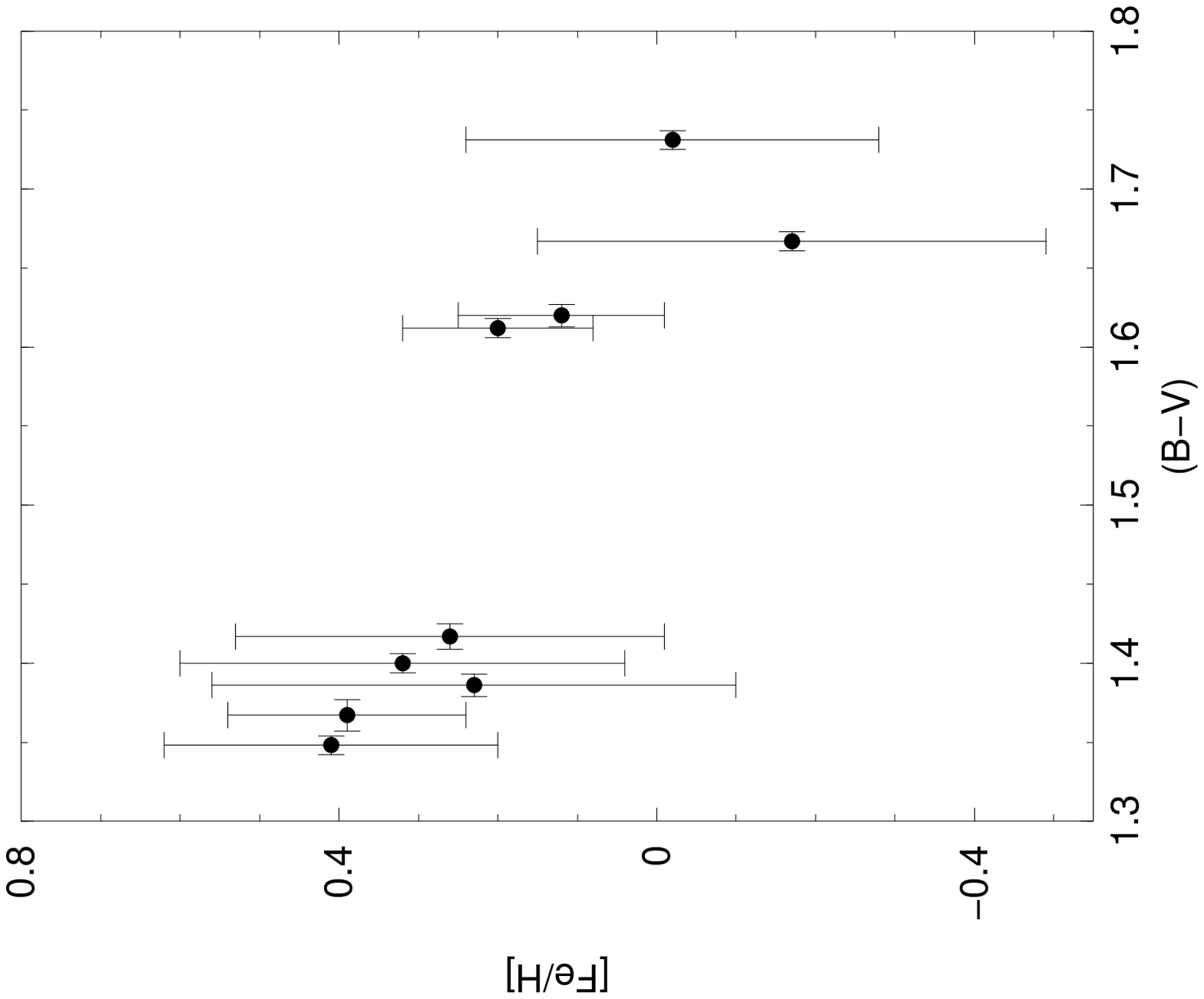,height=15.0cm,angle=270}  }
\caption{Individual \feh\ values from Friel \& Janes as a function of 
$B-V$ color.  Note that the coolest giants give consistently lower
abundances than stars in the hotter group.
}
\label{figfeh}
\end{figure}

In fact, Friel \& Janes' results support a red giant
metallicity somewhat higher than their published value.  Friel \&
Janes' [Fe/H] values for individual stars are plotted against \bv\ in
Fig.\ (\ref{figfeh}), which shows that their sample naturally divides
into two groups.  (Ka\l u\.zny \& Rucinski's (1995) photometry was
used for this plot, though Montgomery \ea's (1994) photometry is
nearly identical).  The four coolest and most blanketed giants give
consistently lower abundances than those in the hotter group.  Friel
\& Janes were keenly aware of the problems associated with such cool
stars: they noted that they couldn't use three of their six indices
for stars with (\bv)$_{\circ} >$ 1.45, due to a non-linear dependence
on color.  Similar problems might affect the remaining three indices
for the coolest stars in a cluster as uniquely metal rich as
NGC\,6791, particularly since calibration standards did not exist for
[Fe/H] $>$ 0 at (\bv)$_{\circ} >$ 1.3 (see Figs.\ 10--12 and 15 in
Friel 1987).  In contrast, the five hotter giants require little or no
color extrapolation from the available high metallicity calibrators,
and thus are likely to be more accurate.  Friel \& Janes' values
of [Fe/H] for the hotter stars are fairly tightly clustered between
+0.25 and +0.40, with a weighted mean of $+0.35 \pm \,0.22$.\footnote{As
discussed by Garnavich {\it et al.}\ (1994) and Tripicco {\it et al.}\
(1995), Friel \& Janes' derived metallicity was also dependent on
their assumed reddening.  However, no further adjustment is required
here, since their assumed value, \ebv\ = 0.12, is quite consistent
with [Fe/H] = +0.4, as shown in \S \ref{sectmodel}.}  This completely
overlaps the range of [Fe/H] found in 2-17 by Peterson \& Green.

The presence of molecular bands might also affect Garnavich {\it et
al.}'s result, obtained by measuring equivalent widths of the Ca~II
infrared triplet lines in very cool giants ($\vi > 1.6$ and $I <
12.5$, equivalent to $\bv \ga 1.55$).  They determined $\feh \approx
+0.22$ by extrapolating Armandroff \& Da\,Costa's (1991) calibration
to metallicities above solar.  Though they were careful to use only
stars with no TiO bands to avoid continuum problems (Olszewski {\it et
al.}\ 1991), CN bands in the same spectral region might cause a
similar, though smaller, nonlinear effect.  NGC\,6791 giants obviously
have much less CN than the carbon star shown in Olszewski {\it et
al.}'s Fig.\ 5, but carbon and especially nitrogen are observed to
have high abundances in 2-17.  Also, blanketing is not the only
concern when using the Ca~II triplet lines to derive metallicity.  As
emphasized by Rutledge, Hesser, \& Stetson (1997), the dependence of
the Ca~II equivalent widths on log~{\it g} and \teff\ for clusters of
different \feh\ and ages is not well understood.  Furthermore, since
it is not clear that the ${\rm [Ca/Fe]}$ enrichment histories were the
same for metal-poor {\it v.s.}\ metal-rich globulars (Carney 1996;
Rutledge \ea\ 1997), a linear extrapolation from halo to greater than
solar metallicities is even more speculative.

Finally, we suggest that extremely high blanketing might also be
responsible for the fact that DDO and Washington photometric indices
(Geisler, Clari\'a, \& Minniti 1991; Piatti, Clari\'a, \& Abadi 1995;
Twarog, Ashman, \& Anthony-Twarog 1997) give consistently lower
abundance estimates for NGC\,6791 than do spectroscopic techniques,
although that lies outside the scope of this paper.  Clearly, the
safest course of action in this unusually high metallicity cluster is
to concentrate on the hottest possible stars, to avoid as much as
possible the effects of blanketing.

In the following section, we take the high dispersion result, $\feh =
+0.4$, as our starting point.  At least some of the light elements
seem to be even further enhanced, but adopting a specific heavy
element distribution is not straightforward.  Peterson \& Green's
results are somewhat ambiguous in the crucial oxygen abundance, and
also the likely distribution depends on the origin of the heavy
element enrichment.  Mixing on the giant branch is expected to be much
less efficient at higher metallicities, but this is very uncertain
without understanding the exact mechanism(s) by which mixing occurs.
Finally, given the substantial age of this cluster, it is possible
that Type I supernovae did not have time to play as large a relative
role as for younger, enriched clusters such as the Hyades.  Due to
these uncertainties, and the lack of suitable color transformation
tables for enhanced [$\alpha$/Fe] values, we were limited to using
scaled solar abundance ratios for all our isochrones.  We will discuss
the probable effects of this input assumption in \S
\ref{alpha}.

This leaves only the question of the assumed helium abundance,
normally not considered to be a source of uncertainty for most open
clusters, because near-solar values of \feh\ imply near-solar $Y$.
Observations of Galactic and extragalactic H~II regions have found
$\dyz = 4\pm 1$ (Pagel \ea\ 1992) and $\dyz = 2$ (Izotov, Thuan \&
Lipovetsky 1997).  An analysis of the spread in the main sequence for
stars of different metallicities lead Pagel \& Portinari (1998) to
conclude that $\dyz = 3\pm 2$.  The large majority of both the stellar
and H~II observations are in metal-poor systems, and so it is not
clear what value of \dyz\ is appropriate for metallicities as high as
in NGC\,6791.  Theoretical determinations of \dyz\ depend strongly on
the lowest initial mass limit needed for the formation of a black hole
(${\rm M_{BH}}$).  For example, Maeder (1992) found $\dyz \sim 1 - 15$
for ${\rm M_{BH}}$ in the range $12 - 120\,\msun$.  Our calibrated
solar models imply $\dyz = 1.7$ (see \S \ref{sectmodel}).

There is little hope of obtaining a reliable independent helium
measurement from the cluster itself.  NGC\,6791 contains several hot,
evolved stars, but these have such high \teff\ and log~{\it g} that
their atmospheric helium abundances must have been significantly
altered from the primordial value by diffusion and/or radiation
pressure, as well as possible mixing on the red giant branch.  In
addition, most of the hot stars are probable binaries (Green \ea\
1999), with the possibility of altered abundances through prior mass
transfer or collisional mixing.  

An alternate technique for determining the helium abundance, the
well-known R method (Iben 1968), has only been calibrated for low
metallicities appropriate for globular clusters (Buzzoni \ea\ 1983;
Caputo \ea\ 1987; Bono \ea\ 1995).  Cassisi \ea\ (1998) have recently
demonstrated that the value of Y derived from such calibrations
depends significantly on the details of the physics included in the
theoretical models, and that the best currently available input
physics imply unrealistic helium abundances for observed globular
cluster R ratios.  One could sidestep the problem of obtaining the
true Y, and simply use existing calibrations (based on earlier input
physics) in order to rank different populations by their helium
abundance, but there are further problems associated with metal-rich
stars in old open clusters.  For example, Renzini (1994) constructed a
modified R$_c$ parameter using photometric data from the OGLE
microlensing survey for Galactic bulge stars in Baade's window
(Paczynski \ea\ 1994).  R$_c$ must be $\le 1.3$ for the bulge stars,
depending on the nature of the field star contamination.  Given $R_c =
0.80$ and $Y = 0.23$ for the globular cluster M3, Renzini determined
that the helium abundance of the metal-rich bulge stars had to lie in
the range 0.30--0.35, corresponding to \dyz\ between 2 and 3.  We can
construct the same R$_c$ for NGC\,6791, using Cudworth's proper motion
sample and Montgomery, Janes \& Phelps (1994) VI photometry.  The
`clump' box contains 30 stars, all known radial velocity members
(Peterson \& Green 1999), while the fainter `RGB' box contains 34
stars whose radial velocity status is unknown.  Thus R$_c$ could be as
low as 0.88, but could only be as high as 1.2 (\dyz = 2) in the
unlikely event that more than 25\% of the high probability RGB members
turn out to be field stars.  Further difficulties include possible
differences in the central concentrations of stars in the clump and
RGB boxes (the proper motion sample only includes stars relatively
near the center of NGC\,6791), and whether nine blue horizontal branch
stars (Green \ea\ 1998) should be added to the clump group for
NGC\,6791 (corresponding stars in the bulge were not considered, and
Paczynski \ea's Fig.~1 is not quite deep enough to show if they
exist).  Even if the calculation of R$_c$ for NGC\,6791 were
straightforward, there is the contrasting case of M67.  The data for
M67 are much less ambiguous, since proper motions and radial
velocities exist for all of the relevant stars out to very large
distances from the cluster center (see \S \ref{sectm67}), and all the
horizontal branch stars are in the red giant clump.  The resulting
R$_c$ ratio for M67 is surprisingly high, $14/10 = 1.40$.  This
contradicts any reasonable scenario in which M67's solar metallicity
and (presumably) solar helium abundance ought to produce a value for
R$_c$ intermediate between M3 on the one hand, and NGC\,6791 and the
bulge on the other hand.

It doesn't help to consider a more classical definition of the R
parameter, specifically R$'$ = the number of clump stars divided by
the number of RGB+AGB stars with luminosities brighter than the clump.
Minniti (1995) found R$'$ to be approximately 1.6 for both his sample
of bulge stars and for several metal-rich disk globular clusters,
whereas R$'$ for a typical metal-poor globular is 1.1.  For NGC\,6791,
R$'$ = 1.37 based on Peterson \& Green's (1999) complete sample
containing 26 clump stars and 19 RGB+AGB stars.  (Or, R$'$ = 1.84 if
the blue horizontal branch stars are included with the clump stars.)
M67 has 7 clump and 9 RGB+AGB stars for a corresponding R$'$ = 0.78.
The relative rankings of NGC\,6791 and M67 are reversed from the
previous case with R$_c$.  Clearly, small number statistics and/or the
underlying metallicity and age dependences make it very difficult to
determine even relative helium abundances in metal-rich old open
clusters using the R method.

Other methods for estimating Y in clusters, {\it e.g.}\ the difference
in magnitude between the turnoff and horizontal branch (Caputo \ea\
1983), are even more poorly suited for use in very metal-rich clusters
or are more model dependent than the R method.  Given the uncertainty
in the true value of the helium abundance in NGC\,6791, we have
elected to consider a range of values from $\dyz = 0$ to 3.

\section{The Stellar Models and Isochrones \label{sectmodel}}

Stellar evolution tracks for masses in the range $0.75 - 1.3 \,\msun$
(in $0.05\,\msun$ increments) were calculated using the Yale stellar
evolution code (Guenther \ea\ 1992).  These models incorporate
the latest available input physics: high temperature opacities from
Iglesias \& Rogers (1996); low temperature opacities from Alexander \&
Ferguson (1994); nuclear reaction rates from Bahcall \& Pinsonneault
(1992) and Bahcall (1989); helium diffusion coefficients from Michaud
\& Proffitt (1993); and an equation of state which includes the
Debye-H\"{u}ckel correction (Guenther \ea\ 1992). Note that this
equation of state yields stellar models which are in good agreement
with those derived using the OPAL equation of state (Rogers 1994;
Chaboyer \& Kim 1995).  The stellar models employ a solar calibrated
mixing length. The models were typically evolved from the zero age
main sequence to the upper giant branch in $\sim 1500$ time steps.  In
each time step, the stellar evolution equations were solved with a
numerical accuracy exceeding 0.01\%.  The models did not include any
overshooting beyond the formal edge of the convection zones.  Main
sequence and turnoff stars in clusters as old as NGC\,6791 do not have
convective cores, and therefore identical ages are obtained with or
without convective overshooting.

All models employed a scaled solar composition (Grevesse \& Noels
1993).  They were calculated for a variety of assumed
metallicities ($\feh = 0.0, +0.3, +0.4, +0.5$) and helium abundances
($\dyz = 0,\, 1,\, 2,\, 3$).  Our calibrated solar model had an
initial solar helium abundance of $Y = 0.264$ and heavy element mass
fraction of $Z = 0.018$.\footnote{Assuming a primordial helium
abundance of $Y = 0.234$ (Olive \ea\ 1997), our calibrated solar model
implies $\dyz = 1.7$.}  Conversion from \feh\ and $\dyz$ to $Z$ and
$Y$ were performed using the standard formulae: 
\begin{equation}
\feh = \log (Z/Z_{\odot}) - \log (X/X_{\odot})
\label{eqzfeh}
\end{equation}
and 
\begin{equation}
Y = Y_{\odot} + (\dyz)(Z - Z_{\odot}).
\end{equation}
 For example, in constructing the $\feh = +0.4$ and $\dyz
= 2$ models, $Z = 0.041$ and $Y = 0.31$ were used.  Note that we have
elected to reference our \dyz\ values to the solar helium abundance,
and not the primordial helium abundance, as there are suggestions that
\dyz\ is not constant for all metallicities.

Isochrones were constructed by interpolating among the evolutionary
tracks using the method of equal evolutionary points (Prather 1976)
for the different \feh\ and $\dyz$ values listed above.  For the fits
to NGC\,6791, the isochrones spanned the age range 6 -- 10 Gyr, in 1
Gyr increments.  The isochrones were transformed from the theoretical
$(\log L,\, \log {\rm T_{eff}})$ plane to observed colors and
magnitudes using color transformations and bolometric corrections
based on the Kurucz (1993) model atmospheres.\footnote{Kurucz's 1995
revisions, described by Bessell \ea\ (1998), do not affect the colors
used here, because the small discontinuities in the 1993 colors for
$\feh \ge 0.0$ only occur at colors bluer than the NGC\,6791 turnoff.}
The transformations used here were kindly supplied to us by Sukyoung
Yi (private communication), who derived them for a wide range of
metallicities (including $\feh = +1.0,\ +0.5,\ +0.3,\ +0.0$), using a
procedure similar to that described by Bessell
\ea\ (1998).  At solar metallicities, the Yi tables agree with Bessell
\ea's Table 1 (from Kurucz's 1995 model atmospheres) to within
$\sim$0.007 mag in \bv\ and $\sim$0.004 in
\vi\ over the range of colors appropriate for old clusters.  Bessell
\ea\ extensively tested their Kurucz-based (ATLAS9) colors for solar
metallicities, and found that all indices agreed extremely well with
observations for $\log {\rm T_{eff}} > 4250$K.  (4250K corresponds to
$\bv \sim 1.35$ at $\feh = +0.4$).

\section{Isochrone Fits \label{sectfits}}

Though NGC\,6791 has been the subject of many photometric studies, for
our purposes the most useful CMD's are those of Ka\l u\.zny \& Udalski
(1992, hereafter KU), Montgomery, Janes, \& Phelps (1994, MJP), and
Ka\l u\.zny \& Rucinski (1995, KR95).  All contain the region of the
cluster covered by Cudworth's preliminary proper motion survey, and
have highly accurate, independently calibrated \bv\ and \vi\ covering
the major sequences from the tip of the giant branch down to well
below the main sequence turnoff.  Using the {\it Acta Astronomica} and
{\it Astronomy \& Astrophysics} electronic datafiles for the KU and
KR95 data, and the MJP data kindly made available by K. Montgomery
(private communication), Green has cross-identified the stars in each
dataset by position, magnitude, and color with those in Cudworth's
(1994) preliminary proper motion survey.\footnote{The
cross-identifications will be available from the Astronomical Data
Center Archive.}  Selecting stars with membership probability $>$ 40\%
defines the major sequences very well.  Since KU's photometry was very
close to that of KR95 in both \bv\ and \vi, we plot only the KR95 and
MJP data in the following sections.

KR95 and MJP each made available two datasets for NGC\,6791, one
covering a wider area of the cluster and the other consisting of
deeper photometry in the cluster center.  For the subset of stars that
are sufficiently well separated from their neighbors to be included in
the proper motion survey, the deeper photometry made a negligible
improvement in the photometric accuracy in both cases.  We used the
data in KR95's Table~2, since it includes all of the proper motion member
stars.  For MJP, both sets gave essentially identical results; in the
following sections we show only the data for the deeper central field.

Isochrones were initially constructed for $\dyz = 2$ and $\feh = +0.3,
+0.4,$ and $+0.5$. For $\feh = +0.4$, isochrones with helium
abundances corresponding to $\dyz = 0,\, 1$, and 3 ($Y = 0.262,\,
0.288,\, = 0.329$ respectively) were also constructed.  In each case,
we determined the best fit to the main sequence and turnoff region --
not the giant branch! -- by requiring that the reddening and distance
modulus be the same for the CMD's in each color.  We assumed E(\vi) =
1.25*(\ebv).\footnote{Dean {\it et al.}\ (1978) show the complete
formula to be E$(\vi) / \ebv\ = 1.25 [ 1 + 0.06(\bv)_0 + 0.14\ebv ]$.
For (\bv)$_0$ appropriate to the main sequence, E(\vi) varies from
1.32 \ebv\ at \ebv\ = 0.05, to 1.33 \ebv\ at \ebv\ = 0.12.  At the
base of the giant branch, E(\vi) are only 0.001--0.003 larger.  For
comparison, we also calculated reddening ratios using the results of
Cardelli {\it et al.}\ (1989).  Quadratic interpolation in their table
3 using the effective filter wavelengths given by Bessell (1990) for a
red (K0 III) star, gives E(\vi) = 1.32 \ebv\ .  The difference in \vi\
between either formula and the expression adopted above is $\leq$
0.01, less than the systematic photometric errors.}  The results are
shown as 2$\times$2 mosaics for each set of input \feh\  and \dyz.
An important constraint to the fits is the fact that there is no
noticeable evidence for a `hook' around the main sequence turnoff.  
Convective cores in higher mass stars produce this hook in isochrones
younger than about 7~Gyr (at these metallicities), resulting in
unacceptable fits to the data.

\begin{figure}
\centerline{\psfig{figure=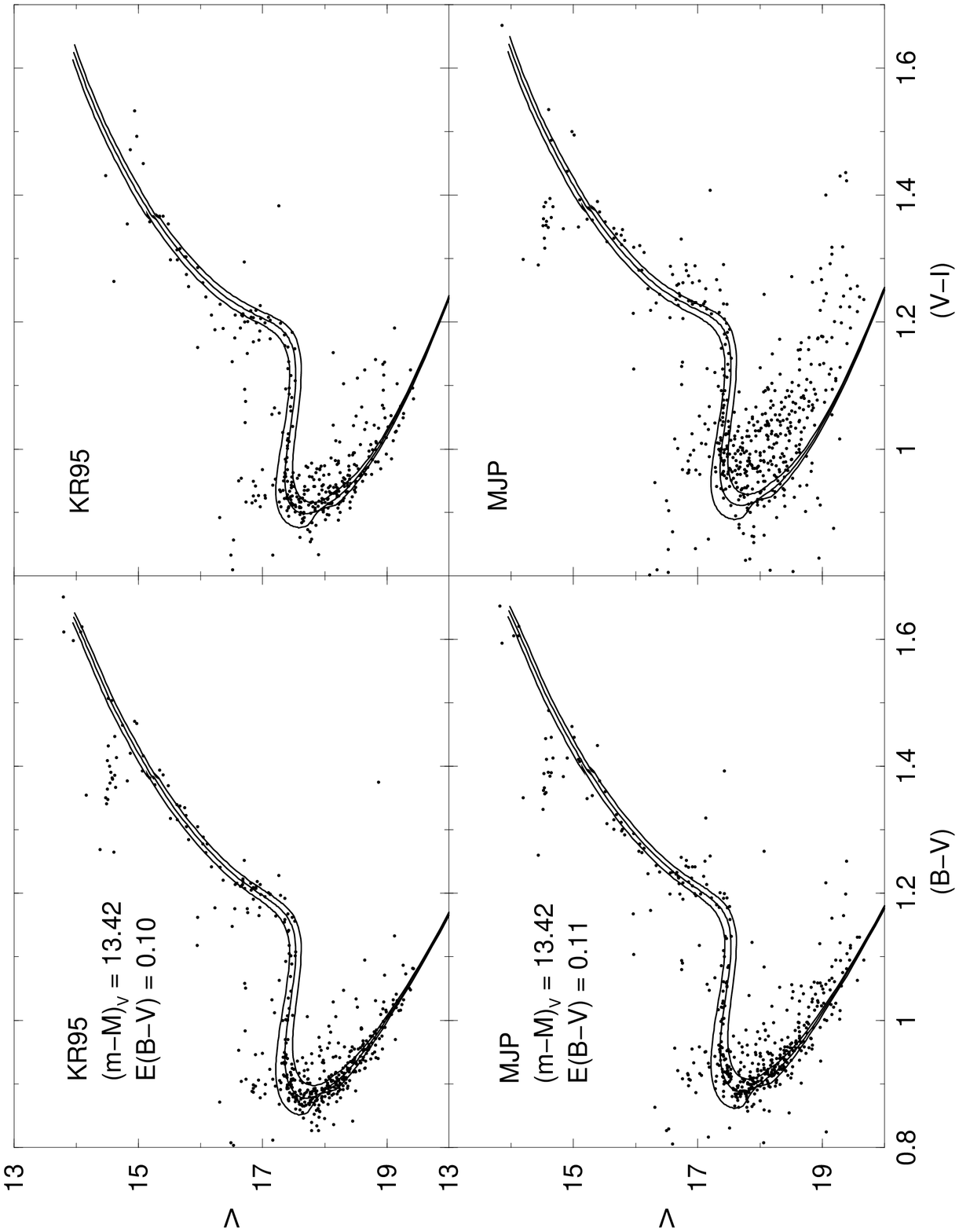,height=19.0cm}  }
\caption{Best simultaneous fit to the \bv\ and \vi\ photometry 
for $\feh = +0.4$, $\dyz = 2$. The fit
to the MJP and KR95 photometry is performed separately.  In each panel,
7, 8 and 9 Gyr isochrones are plotted. 
}
\label{feh04dydz2}
\end{figure}

Fig.\ (\ref{feh04dydz2}) shows the best fit for our adopted $\feh =
+0.4$, with $\dyz = 2$.  The fits are excellent except for the main
sequence $V$ {\it vs.}\ \vi\ data from MJP.  The isochrones and color
calibrations work very well everywhere else, even on the giant
branch where it wasn't at all obvious that this would be the case.
(Note that this implies that the mixing length parameter determined
for the sun appears to work just as well on the giant branch as
for the turnoff of this old cluster.)
Our isochrones fit the KR95 BV data marginally better than the MJP
BV data, but both are quite acceptable.  Our assumed composition and the
recent improvements in the isochrones lead to an age, \ebv, and \dmv\
near the low ends of the previously published ranges.  We will examine
the reasons for this in the following section.

The VI data are puzzling, because the KU, KR95 and MJP data are all
rather close on the giant branch, and the KU and KR95 main sequences
agree very well with each other.  Only the MJP main sequence/turnoff
region is discrepant.  All three groups used the KPNO 0.9m telescope
and standard UBVI filters; KU and MJP used the TEK 512x512 CCD and
Landolt's (1983) calibration standards, whereas KR95 used a TEK
2048x2048 CCD and Landolt's (1992) standards.  The simplest
explanation would be a calibration problem involving MJP's fainter
stars, although we can't see why this should be the case.\footnote{One
can't invoke extinction problems due to Mt.\,Pinatubo, since its
eruption in June 1991 (Grothues \& Gochermann 1992) occurred after the
observations of KU and MJP, and prior to KR95.}  The alternative is
that both the KU and KR95 main sequence colors are in error, in such a
way that the color calibrations are wrong for the \vi\ main sequence,
but correct for the \bv\ main sequence and both giant branches.  The
latter seems counterintuitive since greater difficulties with color
calibrations are normally expected at shorter wavelengths (\bv) and
for cooler stars (the red giant branch).  We will return to this
question in \S \ref{sectm67}.

\begin{figure}
\centerline{\psfig{figure=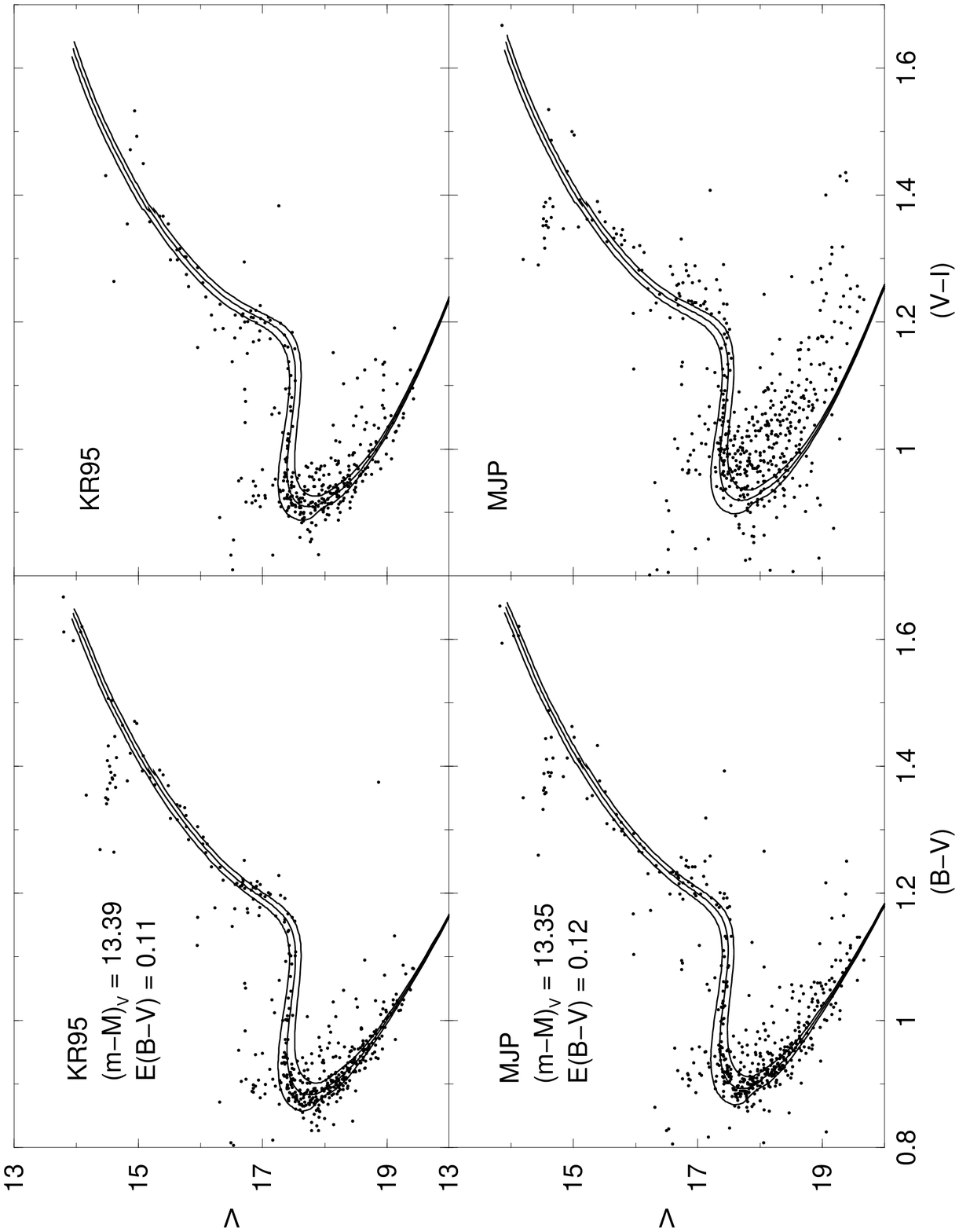,height=19.0cm}  }
\caption{Best simultaneous fit to the \bv\ and \vi\ photometry 
for $\feh = +0.4$, $\dyz = 3$. The fit
to the MJP and KR95 photometry is performed separately.  In each panel,
7, 8 and 9 Gyr isochrones are plotted. 
}
\label{feh04dydz3}
\end{figure}

\begin{figure}
\centerline{\psfig{figure=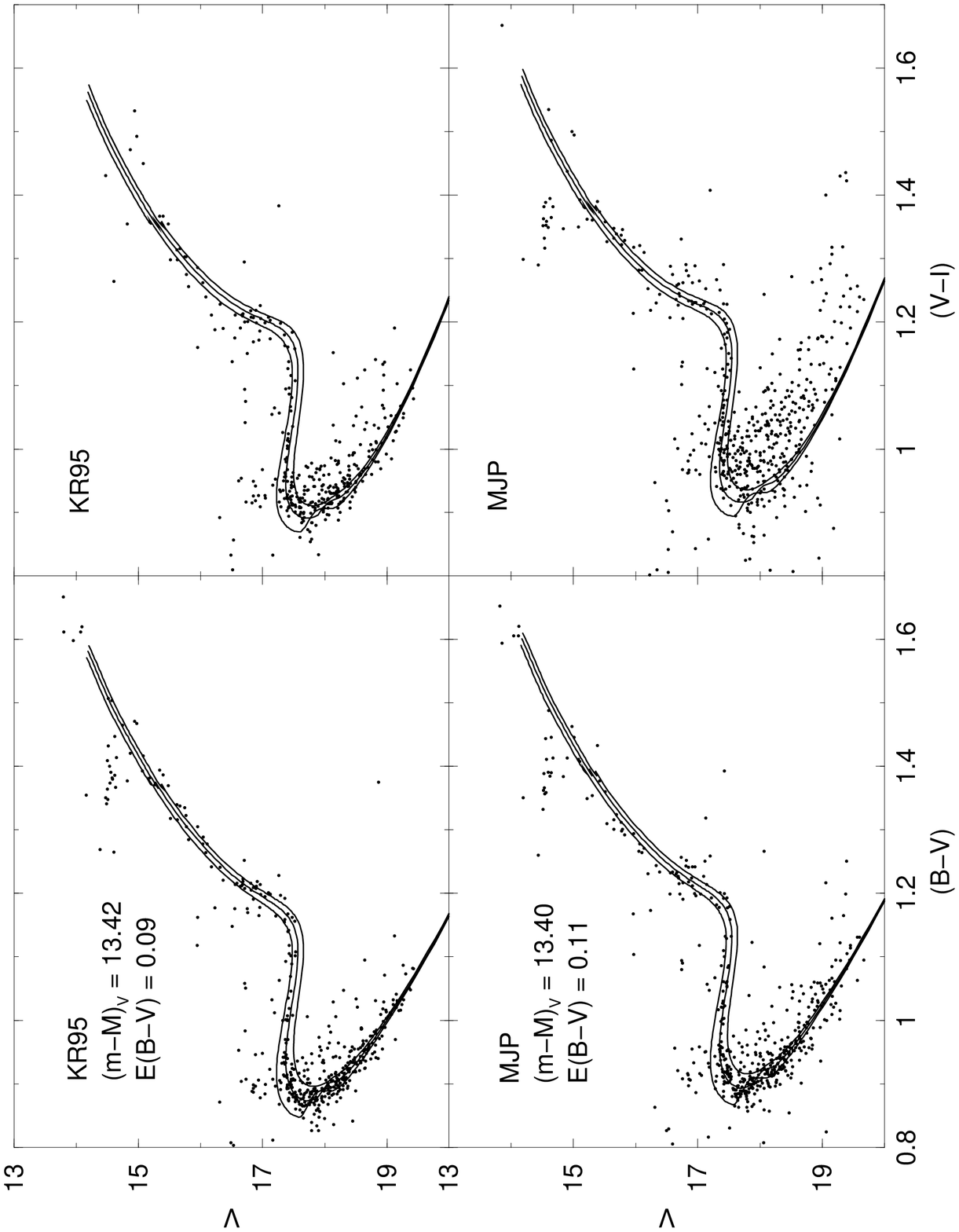,height=19.0cm}  }
\caption{Best simultaneous fit to the \bv\ and \vi\ photometry 
for $\feh = +0.4$, $\dyz = 1$. The fit
to the MJP and KR95 photometry is performed separately.  In each panel,
7, 8 and 9 Gyr isochrones are plotted. 
}
\label{feh04dydz1}
\end{figure}

\begin{figure}
\centerline{\psfig{figure=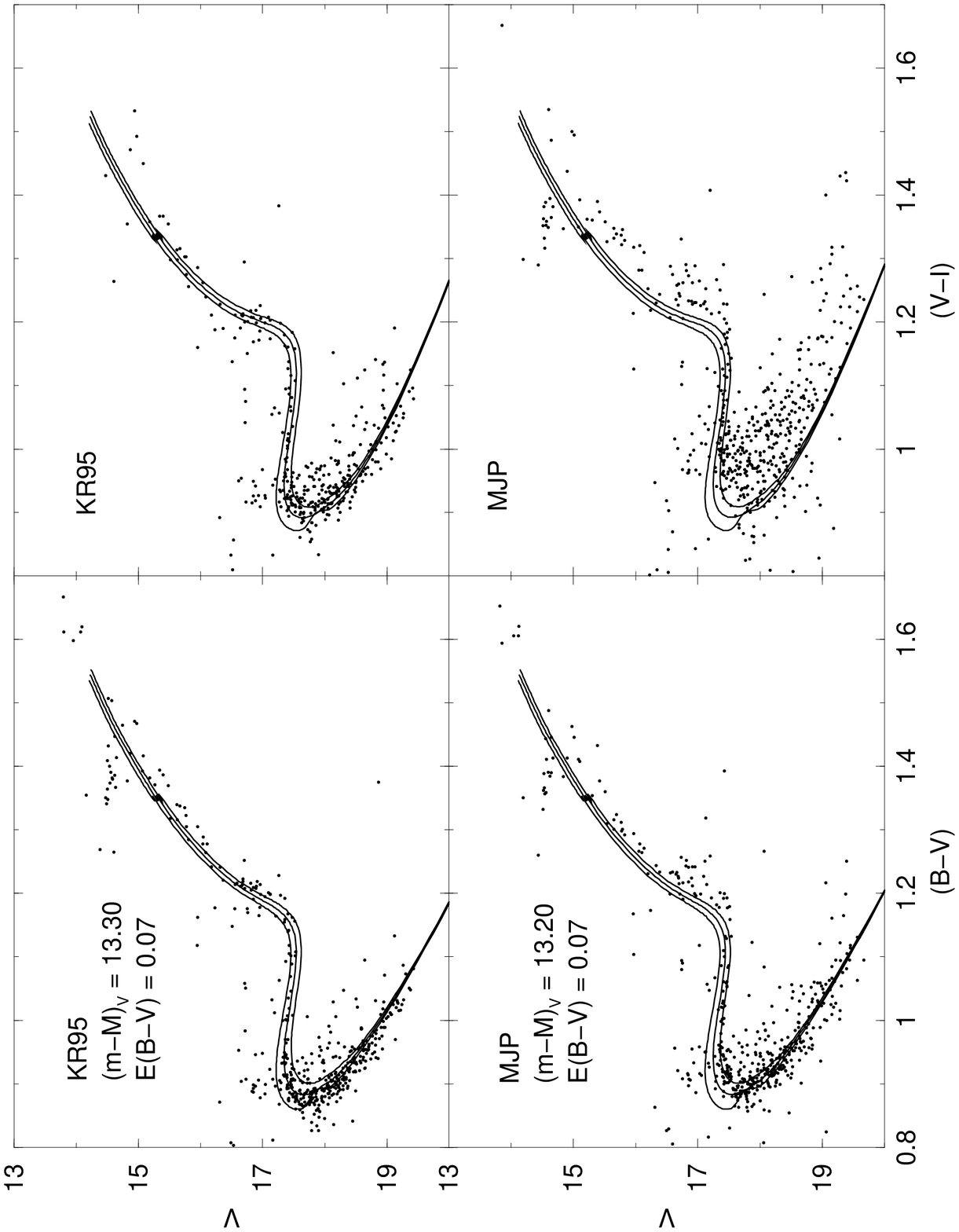,height=19.0cm}  }
\caption{Best simultaneous fit to the \bv\ and \vi\ photometry 
for $\feh = +0.4$, $\dyz = 0$. The fit
to the MJP and KR95 photometry is performed separately.  In each panel,
8, 9  and 10 Gyr isochrones are plotted. 
}
\label{feh04dydz0}
\end{figure}

Figs.\ (\ref{feh04dydz3}) --- (\ref{feh04dydz0}) show the fits for
$\feh = +0.4$ with different values for the helium abundance.
Isochrones for $\dyz = 3$ produce fits that are nearly as good as
those for $\dyz = 2$, whereas those for $\dyz = 0$ show significant
disagreement with the shape of the subgiant branch.  The 8 Gyr $\dyz =
1$ isochrone provides a reasonable fit to the data.  There is a slight
`hook' in this isochrone around the main sequence turnoff, which does
not appear to be present in the data.  However, the present photometry
is not accurate enough to completely rule out the presence of a small
hook, and so we deem the $\dyz = 1$ isochrone an acceptable fit,
though inferior to $\dyz = 2$.  We conclude that the appropriate value
for $\dyz$ lies between 1 and 3 for populations having the same
enrichment history as NGC\,6791.  This is in reasonable agreement with
determinations of \dyz\ based on H~II regions (Izotov \ea\ 1997), and
those based upon the spread of the main sequence in stars of different
metallicities (Pagel \& Portinari 1998).

\begin{figure}
\centerline{\psfig{figure=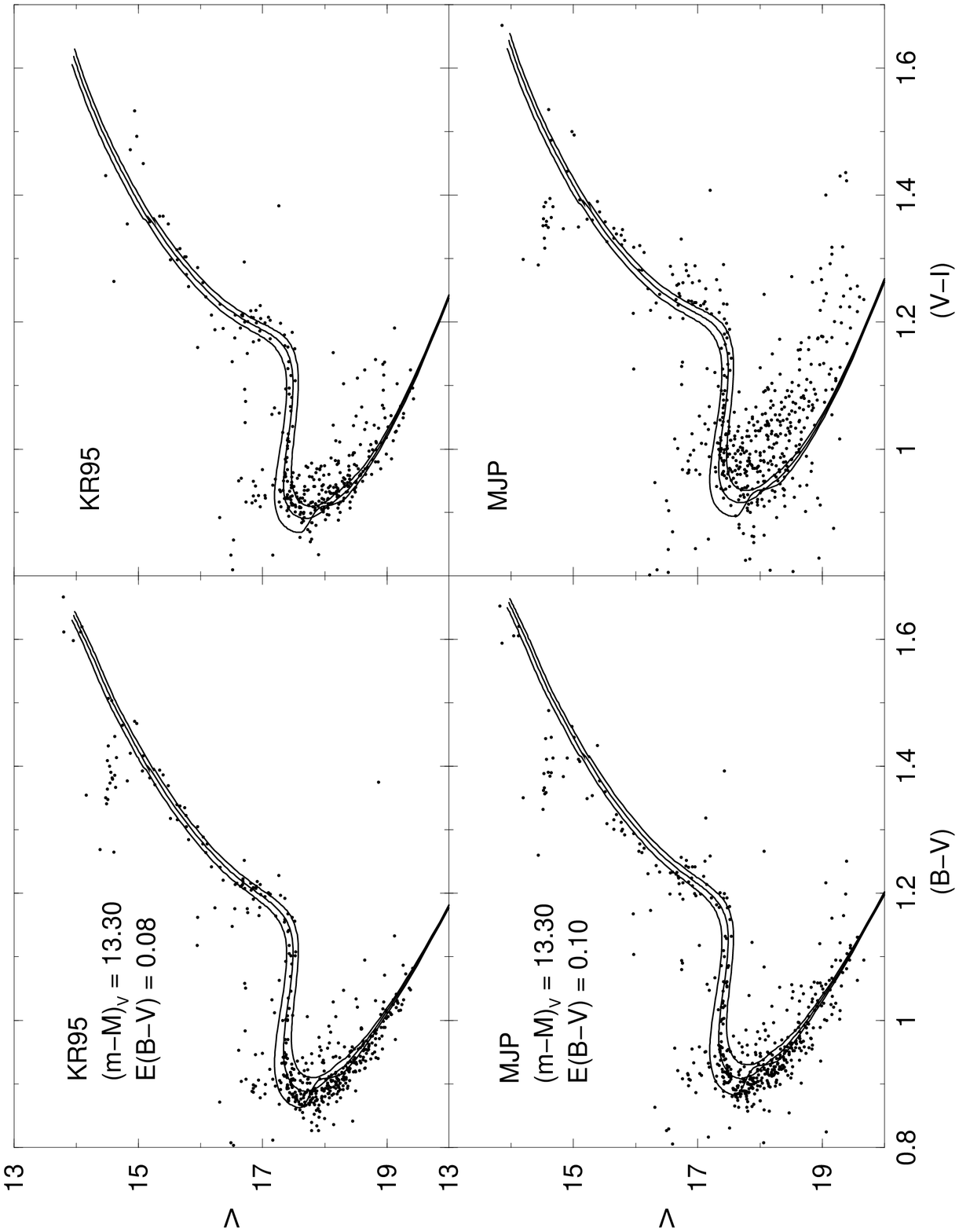,height=19.0cm}  }
\caption{Best simultaneous fit to the \bv\ and \vi\ photometry 
for $\feh = +0.5$, $\dyz = 2$. The fit
to the MJP and KR95 photometry is performed separately.  In each panel,
7, 8 and 9 Gyr isochrones are plotted. 
}
\label{feh05dydz2}
\end{figure}

\begin{figure}
\centerline{\psfig{figure=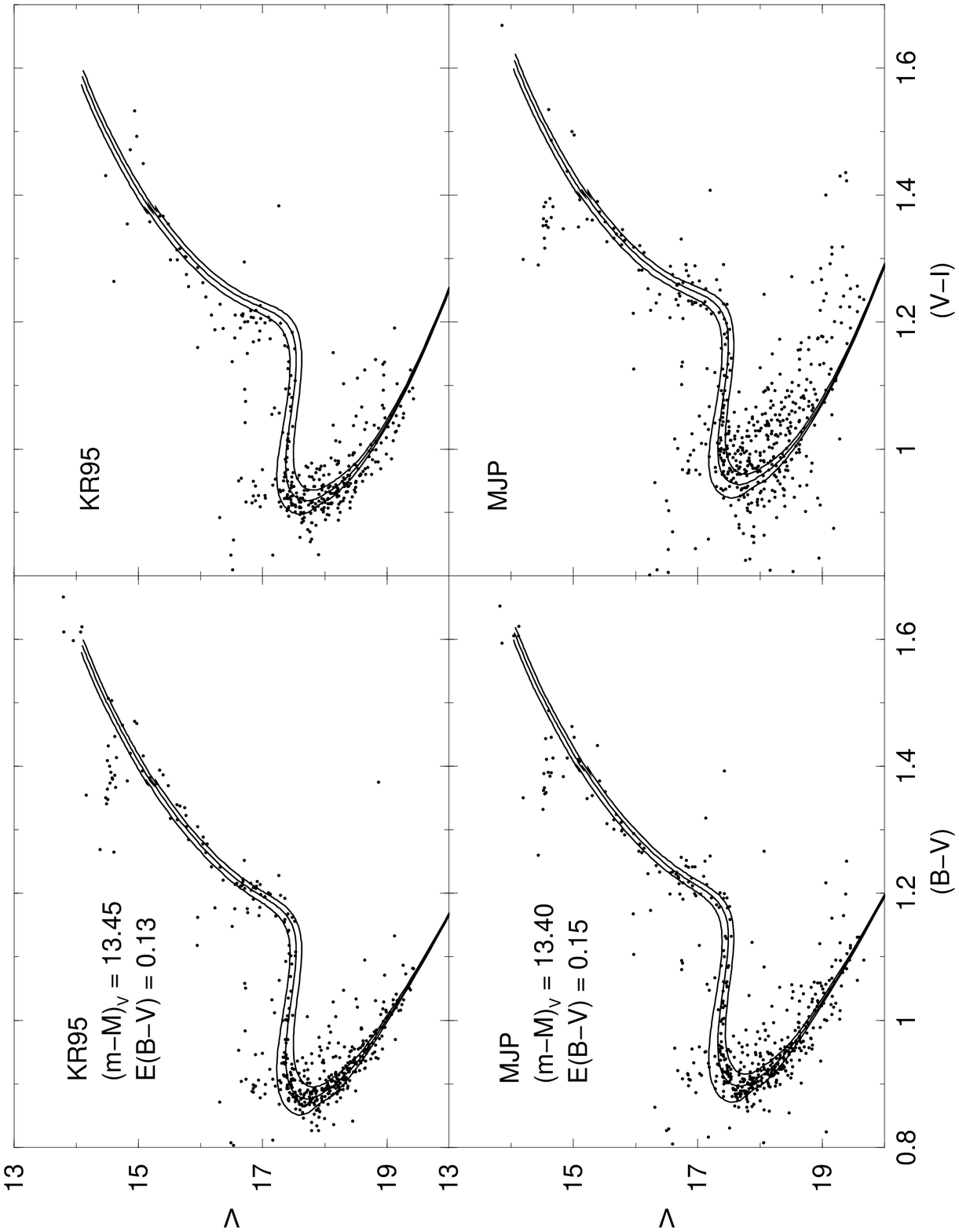,height=19.0cm}  }
\caption{Best simultaneous fit to the \bv\ and \vi\ photometry 
for $\feh = +0.3$, $\dyz = 2$. The fit
to the MJP and KR95 photometry is performed separately.  In each panel,
7,8 and 9 Gyr isochrones are plotted. 
}
\label{feh03dydz2}
\end{figure}

In Figs.\ (\ref{feh05dydz2}) and (\ref{feh03dydz2}), \dyz\ is held
constant and \feh\ is varied.  Neither of these fits are quite as good
as Fig.\ (\ref{feh04dydz2}), and $\feh = +0.5$ is a worse fit than
$\feh = +0.3$, although at this level of detail it is hard to know how
much depends on our assumption of scaled solar abundance ratios.  Taking
$\feh = +0.4$ as the best fit, and assuming that at least some of the
light elements are enhanced, implies that the true \feh\ in NGC\,6791
might be closer to $+0.3$.

\begin{figure}
\centerline{\psfig{figure=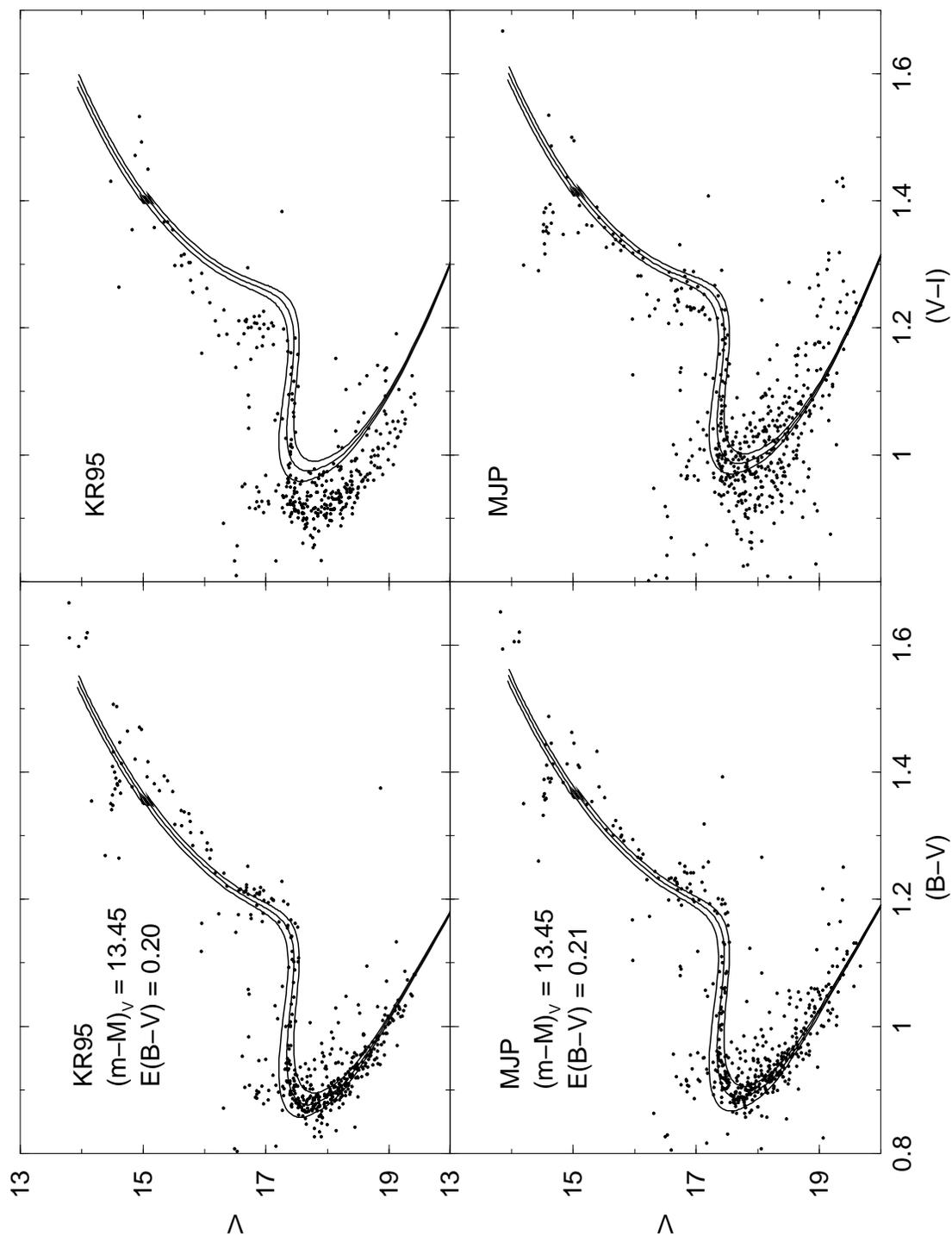,height=19.0cm}  }
\caption{Best fit to the \bv\  photometry using solar metallicity
isochrones.  The corresponding
parameters are used in the \vi\ plot.  The fit
to the MJP and KR95 photometry is performed separately.  In each panel,
8, 9  and 10 Gyr isochrones are plotted. 
}
\label{solar}
\end{figure}

For comparison with some of the previous studies (see the following
section), we also show the best fits for $\feh = 0.0$ in Fig.\
(\ref{solar}). These isochrones were unable to simultaneously match
the \bv\ and \vi\ data, so the fit was optimized for the \bv\ data and
then those same parameters were used in the \vi\ plots.  It is clear
from Fig.\ (\ref{solar}) that using solar metallicity isochrones leads
to rather large reddenings $\ebv \simeq +0.20$.  The derived age of 9
Gyr is 1 Gyr older than that found using our $\feh = +0.4$ isochrones.
The fact that some isochrones are unable to simultaneously match the
\bv\ and \vi\ data demonstrates the power of using multi-color
photometry in performing these fits.  It is interesting to note that
the isochrones predict giant branches that are too blue compared to
the \bv\ data and too red compared to the \vi\ data.

The excellent agreement between our metal-rich isochrones and the KR95
and KU data at all points along the major sequences leads us to
conclude that colors and magnitudes derived using OPAL opacities and
Kurucz's (1993) model atmospheres are very good indeed.  In addition, the
derived cluster parameters are fairly robust to the above variations
in composition.  All of the acceptable fits indicate that $0.08 \le
\ebv \le 0.13$, $13.30 \le \dmv \le 13.45$ and that the age of NGC\,6791 is
$8.0\pm 0.5\,$Gyr. There is a strong correlation between the derived
reddening and distance modulus, with the larger reddening solutions
requiring larger distance moduli.  The heavy element abundance is
consistent with both Peterson \& Green's value for 2-17 and Friel \&
Janes' value for red horizontal branch stars, with a probable $\dyz$ 
between 1 and 3.  Metallicities as low as solar and $\dyz = 0$ are
clearly ruled out.

\section{Discussion, Comparisons \& Tests \label{compare}}

\subsection{The Use of Scaled Solar Abundances \label{alpha}}

For globular cluster metallicities, Salaris, Chieffi \& Straniero
(1993) suggested that the effect of enhancing the $\alpha$ elements by
an average of 0.2\,dex is qualitatively similar to assuming a value
for \feh\ that is larger by about 0.1\,dex.  This suggestion has been
confirmed by other authors (Chaboyer Sarajedini \& Demarque 1992;
Salaris \& Weiss 1998).  If comparable $\alpha$ enhancements are
present in NGC\,6791, our fits might overestimate the metallicity by
this amount.  However, VandenBerg \& Irwin (1997) and Salaris \& Weiss
(1998) showed that this qualitative behavior no longer holds for
metallicities greater than $\feh \sim -0.8$.\footnote{Note that
VandenBerg \& Irwin held \feh\ constant and considered the effects of
additional $\alpha$ enhancement.  Salaris \& Weiss' Fig.\ 4 shows the
alternate point of view, where the total Z is held constant, and
therefore \feh\ is decreased to compensate for increasing $\alpha$
abundances.}  At higher metallicities, the position of the giant
branch becomes insensitive to additional $\alpha$ enhancement
(VandenBerg \& Irwin's Fig.\ 3), while the main sequence and turnoff
continue to be shifted dimmer and cooler.  It is important to note
that the VandenBerg \& Irwin (1997) comparison was done in the
theoretical plane, while Salaris \& Weiss (1998) used the same (scaled
solar) color transformations for their scaled solar and
$\alpha$-enhanced isochrones.  Thus, these studies are unable to
determine how enhancing the $\alpha$ elements affects the appearance
of the isochrones in the observational plane.  

It was the lack of a suitable color transformation for $\alpha$ element
enhanced mixtures which led us to consider only scaled solar
compositions.  It is likely that adopting a non-scaled solar mixture
would have an impact on the calculated isochrones, and the parameters
derived from them (such as the reddening and age of NGC\,6791).  However,
we note that the elemental abundance analysis of Peterson \& Green (1998)
found nearly solar ratios of O and Ca (two of the $\alpha$ elements),
while other $\alpha$ elements (Mg and Si) were enhanced by $0.2 - 0.3$
dex.  Although it can be argued that the presently observed abundance of
O in the evolved star observed by Peterson \& Green (1998) has been
affected by mixing in earlier evolution stages, the same is not true for
Ca.  Thus, the observational evidence for stellar compositions which are
enhanced in the $\alpha$ elements is not clear cut in NGC\,6791.
Enhancing the $\alpha$ elements by a modest 0.2 dex is unlikely to
change the isochrones dramatically. The robustness of the above fits to
variations in \feh\ of $\pm 0.1$ dex implies a similar robustness to
$[\alpha/{\rm Fe}]$ enhancements of order of 0.2 dex.  Detailed models
and more complete observational data will be required to investigate this
further. 

\subsection{Comparison with Previous Isochrone Fits to NGC\,6791 
\label{previous}}

Having concluded that our models fit (most of) the data very well, it
is worth examining the differences between our results and those of
previous investigators.  Table~1 lists recent age determinations for
NGC\,6791.  Most of the ages were determined by fitting isochrones to
the main sequence and turnoff regions in $V$ {\it vs.}\ \bv, except
for Carraro {\it et al.}\ (who found the best match using synthetic
CMD's) and Garnavich {\it et al.}\ (who fit tracks to the giant branch
in $V$ {\it vs.}\ \vi).  In most cases, the tabulated \feh\ and $\dyz$
were those assumed for the isochrones, although some authors argued in
favor of the tabulated compositions after achieving less than
satisfactory results with lower \feh\ (Garnavich {\it et al.}; MJP;
KR95).  Several investigators tried to constrain the range of possible
fits by using independent evidence to set the reddening or distance
modulus.  MJP derived \ebv\ using UBV photometry of foreground field
stars, while KR95 determined a different value from the colors of hot
subdwarf B stars in the cluster.  Garnavich {\it et al.}, Tripicco
{\it et al.}, and KR95 chose their distance moduli based on arguments
about the luminosity of the red giant clump stars.

Fitting technique aside, Table 1 shows that the various investigators
differed in their choice of \feh, helium abundance, opacity tables,
and color transformation tables.  We won't consider further any
results obtained assuming $\feh = 0.0$, in view of the observational
evidence for higher than solar metallicities and also Fig.\
(\ref{solar}) above, but the remaining studies can still be divided
into those who determined ages of 7--8 Gyrs and those who found 9--10
Gyrs.

The effects of different compositions were considered in the previous
section and are clearly too small to be the cause of this difference.
Therefore, we now consider the effects of different opacity and color
transformation tables.  All recent investigations have used either
LAOL or OPAL opacity tables.  Most of the color transformations also
divide into two groups, as well.  Except for Demarque {\it et al.}\
and Meynet {\it et al.}, who used older semi-empirical or empirical
color transformations, the color transformations were derived from
model atmospheres and synthetic colors based on either Kurucz (1993)
or on Bell \& Gustaffson (1978) and Gustaffson \& Bell (1979).  Both
VandenBerg \& Bell (1985) and the MARCS/SSG programs fall into the
latter group.  (Actually Garnavich {\it et al.}\ fall somewhere in
between, since they used a combination of synthetic and semi-empirical
colors.)

\begin{figure}
\centerline{\psfig{figure=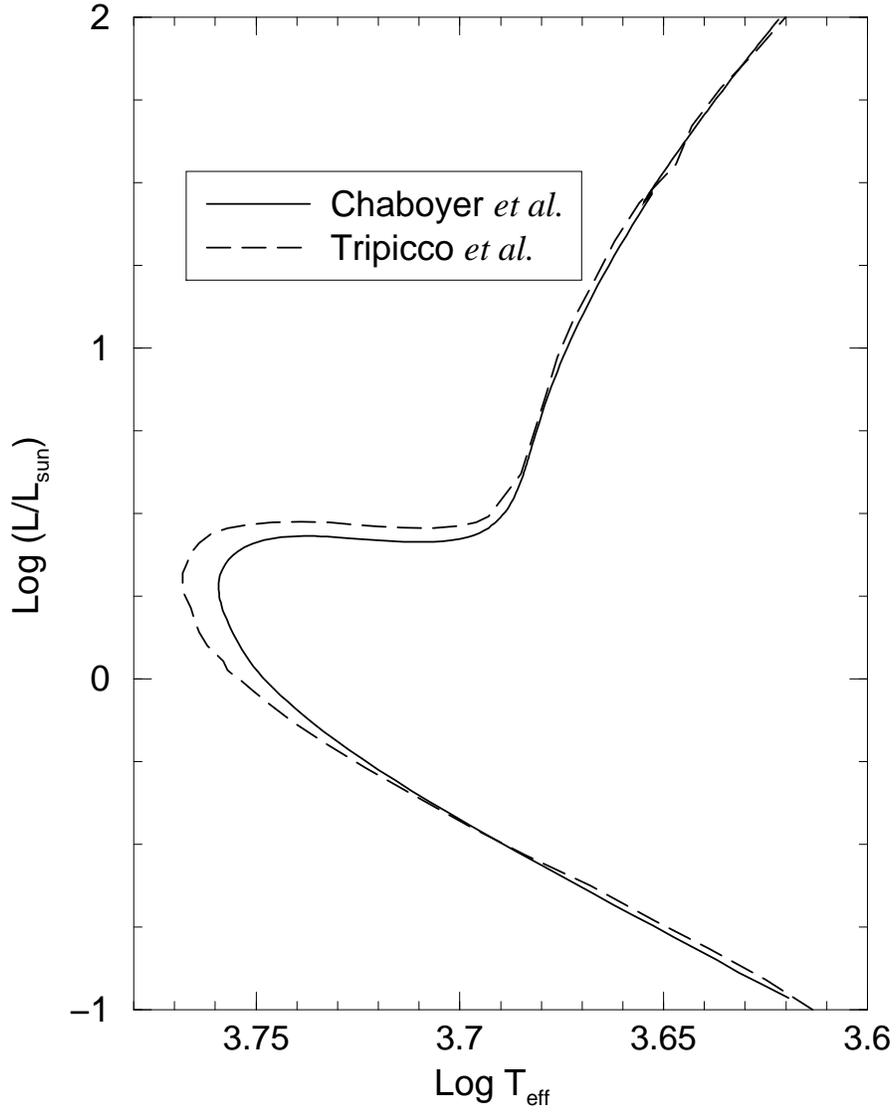,height=15.0cm,angle=270}  }
\caption{A comparison in the theoretical plane between the
isochrones used by Tripicco \ea\ and our isochrones calculated with the
identical composition ($Z = 0.024$, $Y = 0.268$) and mixing length.}
\label{theory}
\end{figure}

M.\ Tripicco kindly supplied us with the isochrones used in their
investigation.  We have plotted their results for $\feh = +0.13$ ($Z =
0.024$), $Y = 0.268$, $\alpha = 1.6$, and 8 Gyr together with our
results for the same parameters.  Fig.\ (\ref{theory}) shows that the
main sequences and red giant branches are nearly identical in the
$\log \teff\ {\it vs.}\ \logl$\ plane, but our isochrones are slightly
cooler by 0.01 in $\log \teff$ at the turnoff.  This is due to the
difference between LAOL and OPAL opacities and the fact that our
models, unlike Tripicco \ea's, include helium diffusion.  The
differences are quite small, as would be expected whenever different
sets of stellar evolutionary models are properly calibrated to the Sun
(Demarque \ea\ 1992).

\begin{figure}
\centerline{\psfig{figure=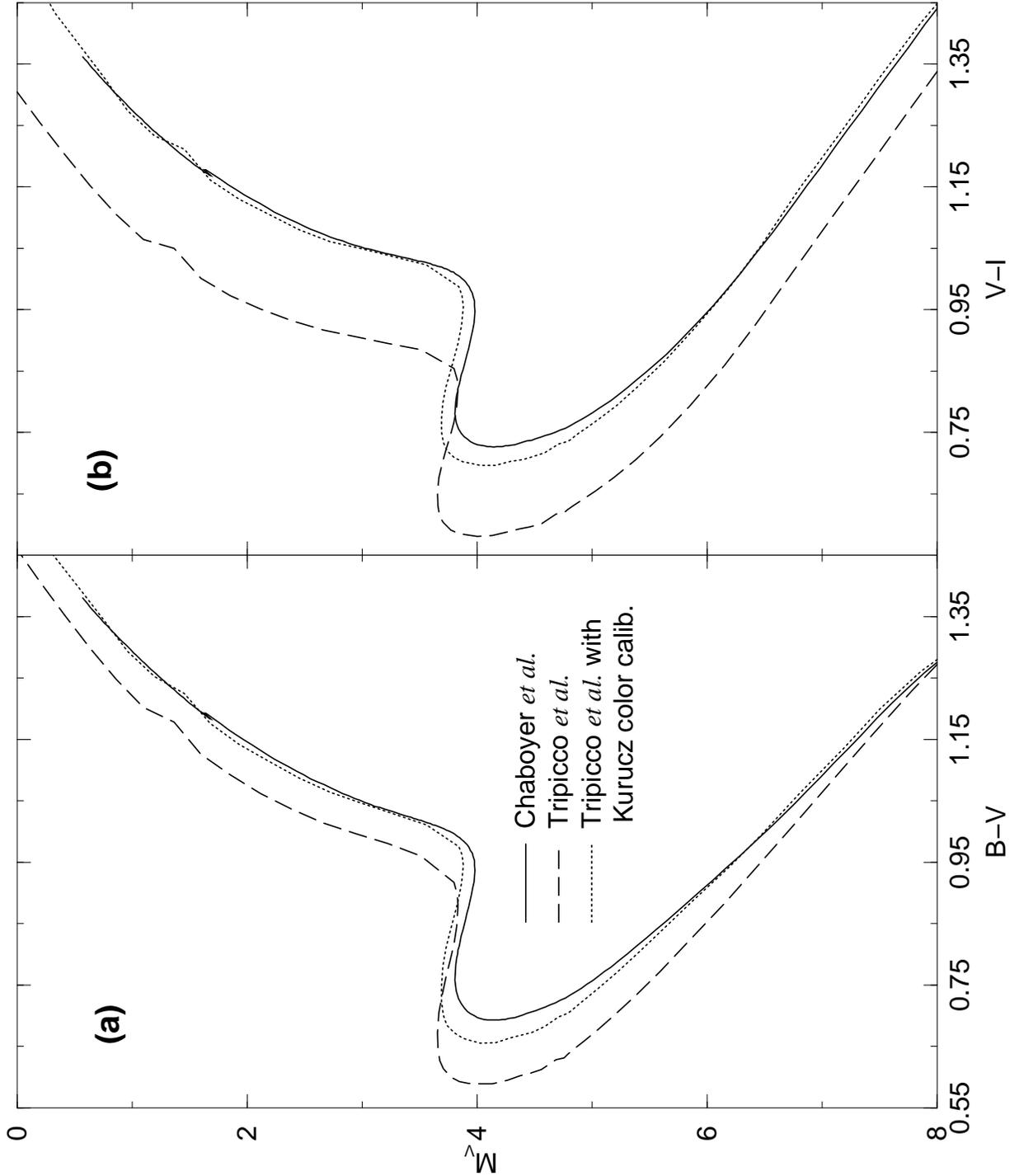,height=19.0cm}  }
\caption{A comparison in the observational plane between the
isochrones used by Tripicco \ea\ and our isochrones calculated with
the identical composition ($Z = 0.024$, $Y = 0.268$) and mixing
length.  To directly compare the effects of the color transformation,
we have additionally converted the Tripicco \ea\ isochrones from the
theoretical plane using the same Kurucz-based color transformation
used for our isochrones.}
\label{tripicco}
\end{figure}

In contrast, the differences in \bv\ and \vi\ between the new
Kurucz-based colors and those based on the older MARCS atmospheres are
substantial for all parts of the color magnitude diagram (Fig.\
\ref{tripicco}).  This illustrates the profound effect of using
synthetic model atmospheres that reproduce real stellar flux
distributions.  We expect that previous isochrone fits to other
clusters using the older synthetic colors will be similarly affected,
although the differences appear to be a function of metallicity.  Reid
(1998) points out the the corresponding consequences for globular
cluster main sequences and turnoffs, comparing D'Antona, Caloi, \&
Mazzitelli's (1997) isochrones, which also used Kurucz (1993) color
transformations, with those of VandenBerg (1985), based on VandenBerg
\& Bell's (1985) colors.  As discussed by Bessell {\it et al.}\ 
(1998), the new NMARCS grid results in colors that match the
observations as well as Kurucz's (ATLAS9) colors.  Therefore, we
expect that isochrones using NMARCS would result in lower $\ebv$ for
NGC\,6791 and good fits for ages of about 8 Gyr, in agreement with
Carraro {\it et al.}, Ka\l u\.zny \& Rucinski, and this paper
(assuming they also chose a greater than solar helium abundance
consistent with the adopted \feh).

Two of the studies listed in Table~1 used the same opacities and color
transformations as we did, and nearly identical \dyz, so the
differences between our results and theirs are more subtle.  Carraro
\ea\ derived an identical age and reddening with a significantly 
lower quoted value for \feh.  We would have expected our ages to
agree, but their \ebv\ to be higher than ours.  We note that they
derive lower reddenings than the canonical values for other old open
clusters as well: for M67, NGC\,2243, Be\,39, NGC\,188, and NGC\,6791,
they find 0.01, 0.01, 0.10, 0.03, and 0.10, respectively.  By
comparison, Twarog, Ashman, \& Anthony-Twarog (1997) derive 0.04,
0.06, 0.12, 0.10, and 0.16 from homogeneous DDO photometry of these
clusters.  Carraro \ea\ 's somewhat smaller \ebv\ might be a result of
their synthetic CMD fitting technique, which makes use of fits to the
red giant branch (and clump), in addition to the main sequence and
turnoff.  The input physics used in constructing the stellar models
employed by Carraro \ea\ differed somewhat from our input physics (\S
\ref{sectmodel}).  In particular, their models did not include
diffusion, used low temperature opacities from LAOL, and used older
values for the $pp$ and CNO nuclear reaction rates.  The effects of
these differences in input physics should be small, except on the giant
branch.  Finally, Carraro \ea\ used a different relation between \feh\
and $Z$.  The composition they adopted for NGC\,6791, $Y=0.305$ and
$Z=0.030$, corresponds to $\feh = 0.17$, using their Eq.\ (1).  The
same Y and Z in our Eq.\ (\ref{eqzfeh}) gives $\feh = 0.255$.

\begin{figure}
\centerline{\psfig{figure=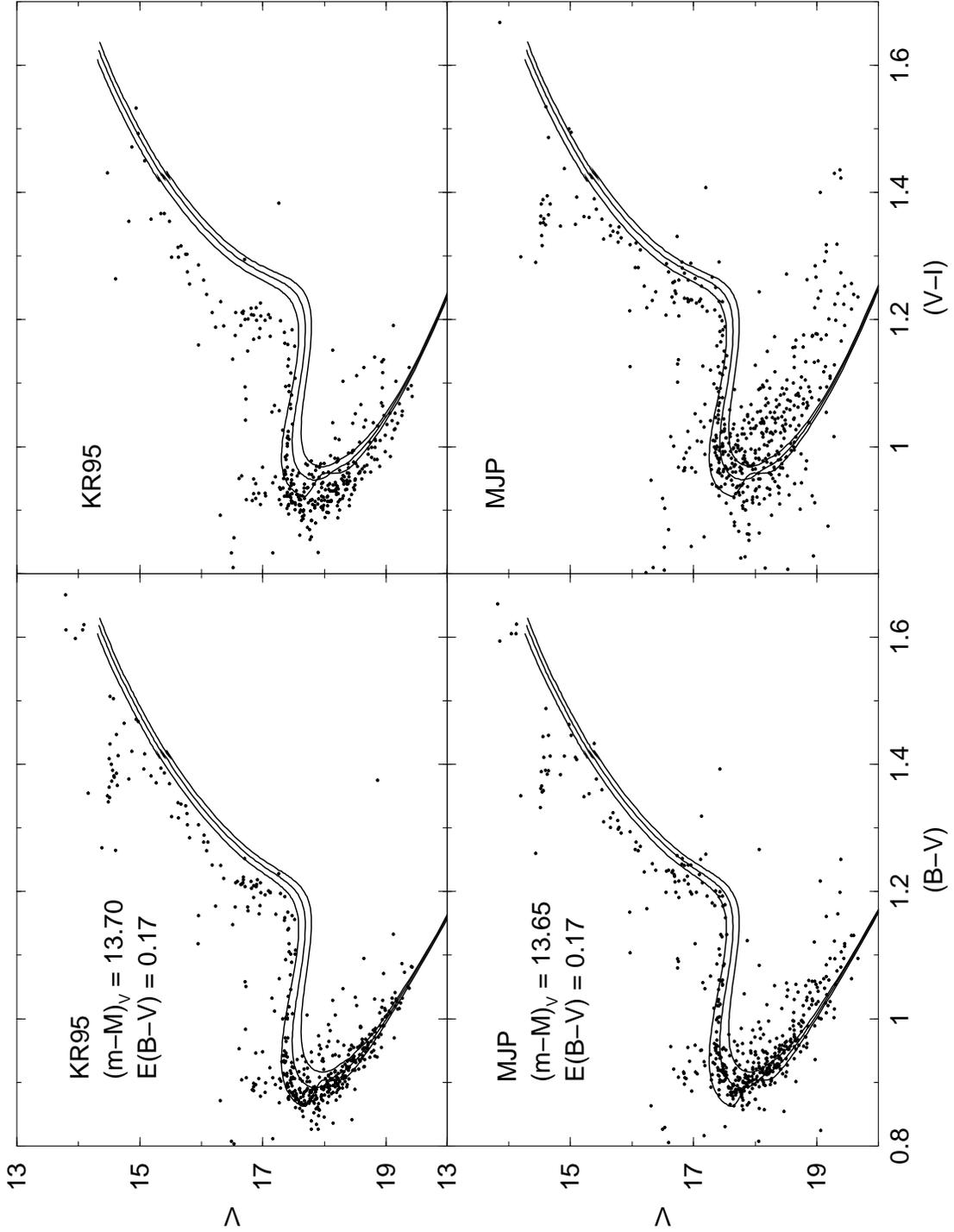,height=19.0cm}  }
\caption{Best fit to the \bv\  photometry using a reddening of $\ebv =
+0.17$ and $\feh = +0.3$ (values assumed by KR95).  The corresponding
parameters are used in the \vi\ plot.  The fit to the MJP and KR95
photometry is performed separately.  In each panel, 6, 7 and 8 Gyr
isochrones are plotted.  }
\label{highred}
\end{figure}

KR95 used the same Bertelli \ea\ (1994) isochrones as Carraro \ea\
(1994), so the differences between their isochrone results and ours
should be similarly small.  The main difference between the present
work and KR95 is that they were not able to interpolate between the
theoretical isochrones for $Z = 0.02$ and $Z = 0.50$, so they had to
use a variety of elegant arguments concerning the interrelation of
\feh, \ebv, and \dmv\ in order to derive their age of 7.2 Gyr.  Most 
importantly, they adopted $\ebv = 0.17$ based on the difference
between the observed and theoretically predicted colors for subdwarf B
stars, and suggest that MJP's $\ebv = 0.10$ from foreground dwarfs was
merely a lower limit to the true reddening.  We believe that the
theoretical subdwarf B colors, which Liebert \ea\ (1994) calculated
using pure hydrogen model atmospheres (Wesemael \ea\ 1980), are
probably several hundredths of a magnitude too blue, and that KR95's
derived reddening is therefore an overestimate.  When we adopt $\ebv =
0.17$ and $\feh = +0.3$, we get the isochrone fits shown in Fig.\
(\ref{highred}).  Note that a reasonable fit to the data cannot be
found for this reddening.  In particular, a good match to the main
sequence requires a substantially higher distance modulus, which
implies a younger age (6 Gyr) for the cluster.  However, the 6 Gyr
isoschrone has a hook around the main sequence turnoff which is not
seen in the data.

Our value of $0.10$ for \ebv\ can only be considered accurate if all
aspects of the stellar models, including the color transformations,
are exactly correct.  The Kurucz models we are using predict $\bv =
0.67$ for the Sun.  The actual solar \bv\ has been somewhat
controversial (VandenBerg \& Poll 1989), but recent work has converged
on $\bv_\odot = 0.65$ (Bessell \ea\ 1998, appendix C).  This would
seem to indicate that the Kurucz (1993) models are slightly too red,
though such a correction wouldn't necessarily apply at higher
metallicities (in particular, note \S \ref{secthyad}).  
A possible compensating color error might result from our models being
slightly too blue on the main sequence and turnoff, due our use of
scaled solar abundances, but this is very uncertain.  We don't know
the true abundance pattern in unevolved NGC\,6791 stars, and, in
any case, there are no color transformations available for
alpha-enhanced mixtures.

\subsection{Comparison with the Hyades ZAMS \label{secthyad}}

\begin{figure}
\centerline{\psfig{figure=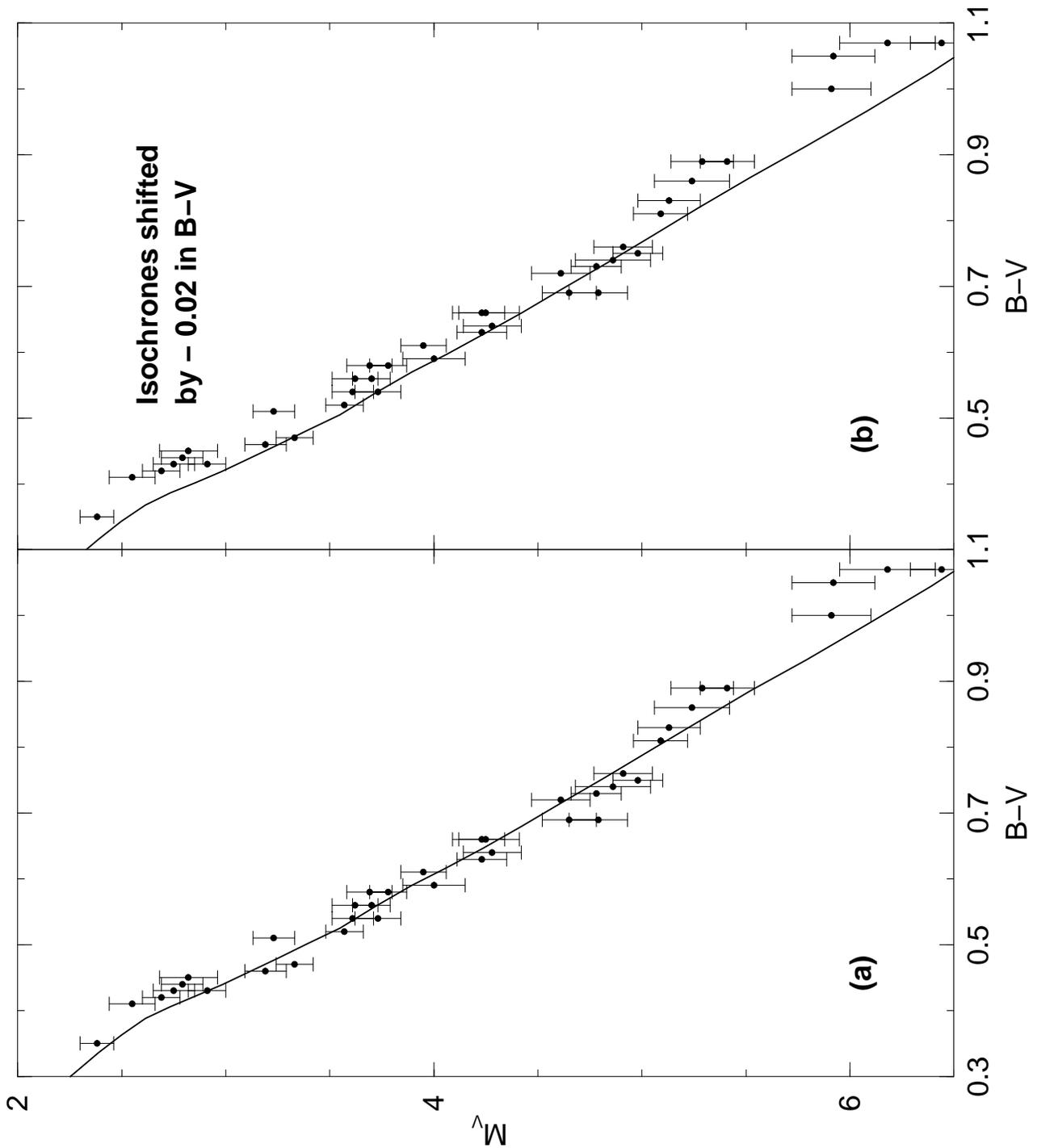,height=19.0cm}  }
\caption{The theoretical ZAMS for $\feh = +0.13$ and $\dyz = 2$ 
is compared to the observed ZAMS of the Hyades, as inferred by
Hipparcos (Perryman \ea\ (1997). As shown in the left panel, the
isochrones provide an excellent fit to the observed ZAMS, without any
color or magnitude shifts.  Indeed, as shown in the right panel,
shifts as small as $0.02$ in \bv\ lead to noticeably inferior fits to
the data.}
\label{hyad}
\end{figure}

In fitting the isochrones to NGC\,6791, we relied extensively on the
location of the unevolved main sequence to determine the distance
modulus and reddening.  The derived parameters are only as
reliable as the location of our theoretical main sequence.  Our
predicted location for the metal-rich unevolved main sequence may be
tested by comparing our isochrones to the unevolved main sequence in
the Hyades ($\feh = +0.13$, Boesgaard \& Friel 1990), as determined by
Hipparcos (Perryman \ea\ 1998).  All of the single, unevolved stars
with good \bv\ photometry and parallaxes with errors less than 10\%
were selected from the list presented by Perryman \ea.  The
parallaxes and photometry were used to derive the absolute $V$
magnitude of the stars.  These stars (with associated errors) are
plotted in Fig.\ (\ref{hyad}).  Our $\feh = +0.13$, $\dyz = 2$
isochrone is overplotted, with {\it no} adjustable parameters.
It is clear from this figure that the unevolved main sequence in our
metal-rich isochrones is in excellent agreement with the observations.
As shown in Fig.\ (\ref{hyad}b), shifting the isochrones by as little
as 0.02 in \bv\ results in a noticeably inferior match to the data.
This test gives us some confidence that our derived distance moduli
and reddening for NGC\,6791 should be reliable.  Perryman \ea\ 
came to a similar conclusion for Hyades stars, namely that modern
stellar isochrones provide a good match to the unevolved main sequence
in the Hyades.

\subsection{Comparison with M67 \label{sectm67}}

\begin{figure}
\centerline{\psfig{figure=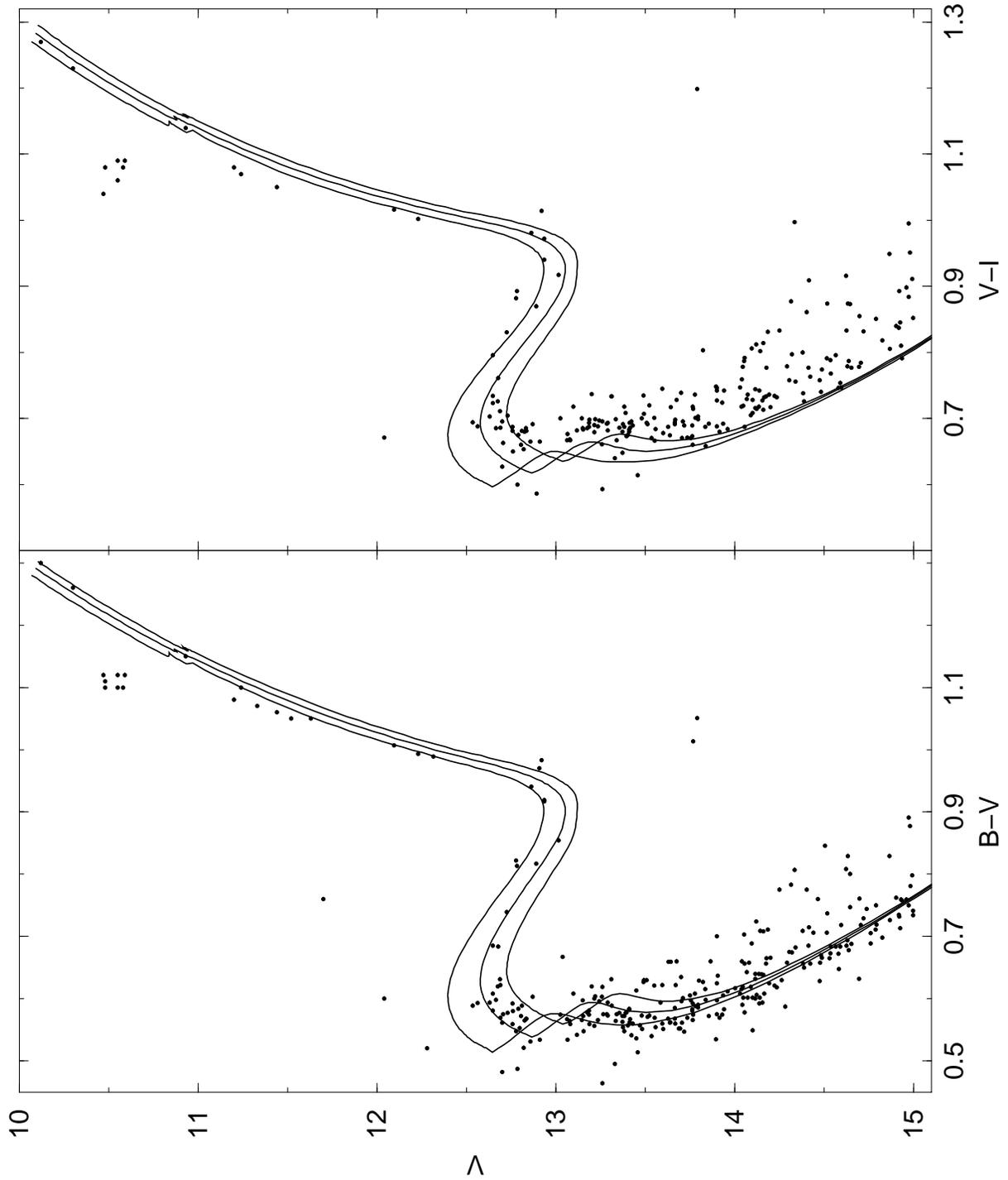,height=19.0cm}  }
\caption{Best simultaneous fit to the \bv\ and \vi\ photometry of M67
(Montgomery \ea). In each panel, 3, 3.5 and 4  Gyr isochrones are
plotted assuming $\dmv = 9.70$ and $\ebv = 0.02$.
}
\label{m67}
\end{figure}

We have seen how well our isochrones match the shape of NGC\,6791 and
the distance modulus of the Hyades, both of which have higher than
solar metallicity.  Bessell {\it et al.}'s (1998) thorough
investigation of colors based on Kurucz's and other new synthetic
atmospheres were only tested with stars of approximately solar
composition.  The well-studied, bright, solar metallicity cluster M67
allows us to compare our isochrones in this same regime.  The
reddening in M67 has been shown by numerous investigators to be $0.02
< \ebv < 0.05$ ({\it e.g.}\ Nissen, Twarog, \& Crawford 1987).  We fit
isochrones for [Fe/H] = 0.0 ({\it e.g.}\ Twarog \ea\ 1997) and the
solar helium abundance to the BV and VI photometry of Montgomery,
Marschall, \& Janes (1993), in the same way that we did for NGC\,6791.
Field stars were eliminated by cross-identifying Montgomery {\it et
al.}'s (1993) stars with those in Girard {\it et al.}'s (1989) proper
motion survey and with Mathieu {\it et al.}'s (1986) radial velocity
survey.  We further removed all known binaries (Mathieu {\it et al.}\
1986; Mathieu, Latham, \& Griffin 1990) in order to clarify the
position of the major sequences, particularly near the turnoff.  The
results are shown in Fig.\ (\ref{m67}).

For reasonable values of \ebv, we find exactly the same behavior we saw
before with MJP's NGC\,6791 data, namely an excellent fit in BV
and increasingly discrepant behavior on the main sequence in VI.
Since the color transformations cannot be this much in error at solar
metallicity, and our isochrones correctly reproduce the BV main
sequence, we are forced to conclude that there are errors in the main
sequence VI photometry of both Montgomery, Marschall, \& Janes and
MJP.  (We note that they observed stars in both M67 and NGC\,6791
during some of the same observing runs.)

Our derived age of 3.5 Gyr, our reddening, $\ebv = 0.02$, and
resulting distance modulus, $\dmv\ = 9.70$, are in reasonable
agreement with previous results for M67.  Carraro \ea\ (1996) found a
good fit with an age of 4.0 Gyr, $\ebv = 0.025$ and $\dmv = 9.65$.
Dinescu \ea\ (1995) found $\ebv = 0.03 - 0.06$ and $\dmv = 9.7 - 9.8$,
with an age of $4.0\pm 0.5$ Gyr.  Meynet \ea\ (1993) determined an age
of 4 Gyr, $\ebv = 0.03$ and $\dmv = 9.6$.  In their photometry paper,
Montgomery \ea\ (1993) favoured an age of $4-5$ Gyr, $\ebv = 0.05$ and
$\dmv = 9.5$ based upon the isochrones of Castellani \ea\ (1992) and
VandenBerg (1985), but noted that these isochrones did {\it not}
provide a good match to the data.  Tripicco, Dorman, \& Bell (1993)
found an age of 5.0 Gyr, with $\ebv = 0.032$ and $\dmv = 9.55$, using
the same isochrones and color transformations in M67 that Tripicco
\ea\ (1995) used for NGC\,6791.

\section{Summary and Implications} 

Using a new grid of scaled solar isochrones incorporating the most
recent input physics and color transformations, we conclude the
following:

(1) The high value of $\feh = +0.4$ found by Peterson \& Green (1998) for
a blue horizontal branch star in NGC\,6791 is very consistent with
fits to the new isochrones presented here, using color transformations
based on Kurucz's (1993) model atmospheres.  Our isochrones fit the
shape of the observed data sufficiently well along the major sequences
of three independent $V~{\it v.s.}\ \bv$ CMD's and two $V~{\it v.s.}\
\vi$ CMD's that we are able to set useful limits on \feh: $+0.4 \pm
0.1$.

(2) The above fits similarly constrain \dyz\ to be $2 \pm 1$.
In particular, a value of zero is clearly ruled out.  

(3) The recent Kurucz-based color transformations are a big improvement
over earlier transformations for metal-rich compositions.  Bessell
\ea\ (1998) have already demonstrated how well they reproduce the
colors of stars with approximately solar composition.  We show here
that they reproduce not only the shape of the NGC\,6791 turnoff and
subgiant branch, but also the giant branch, where we had no {\it a
priori} reason to expect good agreement at such a high metallicity.
Most convincingly, for a nearby metal-rich cluster having a
well-determined distance (from Hipparcos) and reddening, we find an
excellent fit to the Hyades zero age main sequence, with no adjustable
parameters.

(4) Our derived reddening for NGC\,6791, $\ebv \sim 0.10$, which
depends primarily on the assumed composition and the color
transformation, falls at the low end of the previously determined
range.  Higher values for the reddening found in some previous
investigations are a general consequence of using older color
transformations, and/or significantly underestimating the metallicity.

(5) The age of NGC\,6791 is robustly determined to be 8.0~Gyr with an
internal error of $\pm\ 0.5$~Gyr.  Reasonable variations in either
\feh\ or \dyz\ appear to have only a small effect on the derived age.
The external error depends on both the input physics and the color
transformation.  Large changes in the input physics, {\it e.g.}\
including diffusion and using OPAL instead of LAOL opacities, can
affect the resulting ages by up to $\sim 1$ Gyr, as demonstrated by
the comparison between our 8 Gyr isochrone and that of Tripicco \ea
(1995).  The choice of color transformation is potentially an even
greater factor.  However, since any future improvements to the
opacities and synthetic atmospheres will necessarily have to reproduce
the current excellent agreement with the observed colors of both field
stars and the Hyades, future isochrones should not differ from the
current versions by nearly as much as did previous isochrones
utilizing older input physics and color transformations.  Therefore,
the derived age of 8~Gyr for NGC\,6791 is likely to be accurate to
better than 1~Gyr.


We conclude with a few implications of these results: The moderately
`young' age of 8~Gyr $\pm 1$ found here in no way changes the relative
age rankings of old clusters (Phelps \ea\ 1994; Janes \& Phelps 1994).
It only maintains the age ranking of NGC\,6791, and is consistent with
the reduced ages now being derived for all clusters as a result of
comparisons with the newest isochrones, color calibrations, and
Hipparcos distances ({\it e.g.}\ Gratton \ea\ 1997; Reid 1998;
Chaboyer \ea\ 1998; Cassisi \ea\ 1998; Grundahl, VandenBerg, \&
Andersen 1998; Salaris \& Weiss 1998).  The confirmation of an
enhanced metallicity in NGC\,6791 reinforces the potential role of
this cluster as a resolved proxy for old, metal-rich extragalactic
populations, such as the metal-rich bulges of elliptical galaxies.

\acknowledgments {We would like to thank Mike Tripicco for sending us 
the isochrones used by Tripicco {\it et al.}, making it possible to
separate the effects of opacities and color transformations, Sukyoung
Yi, for generously providing us with an improved table of colors based
on Kurucz's models, and Kent Montgomery and Ken Janes, for sending us
their disk data files for NGC\,6791 and M67.  We also thank Eileen
Friel and Ruth Peterson for their constructive comments and insights,
and Mike Bessell for further information concerning the color
transformations.

BC was supported for this work by NASA through Hubble Fellowship grant
number HF--01080.01--96A awarded by the Space Telescope Science
Institute, which is operated by the Association of Universities for
Research in Astronomy, Inc., for NASA under contract NAS 5--26555.
EMG and JL were partially supported by the National Science Foundation
under grant AST-9731655.}

\clearpage

\end{document}